\documentclass[pra,twocolumn, preprintnumbers,amsmath,amssymb,nofootinbib, accepted=2018-09-13]{quantumarticle}
\usepackage{graphicx, color, graphpap}% Include figure files
\usepackage{amsmath}
\usepackage{enumitem}
\usepackage{amssymb}
\usepackage{amsthm}
\usepackage{pstricks}
\usepackage{float}
\usepackage{multirow}
\usepackage{hyperref}
\usepackage{overpic}
\usepackage[T1]{fontenc}

\newcommand{\bra}[1]{\langle {#1} \vert}
\newcommand{\ket}[1]{\vert {#1} \rangle}

\usepackage{hyperref}
\usepackage{cleveref} % must be loaded after hyperref

\begin{document}

\title{Distinguishability times and asymmetry monotone-based quantum speed limits in the Bloch ball}
%\date{\today}
\author{T.J. Volkoff}
%\orcid{0000-0002-5511-3913}
\affiliation{Department of Physics, Konkuk University, Seoul 05029, Korea}
\email{volkoff@konkuk.ac.kr}
\author{K.B. Whaley}
%\homepage{http://quantum-journal.org}
%\orcid{0000-0002-5511-3913}
\affiliation{Berkeley Quantum Information and Computation Center and Department of Chemistry, UC Berkeley, Berkeley, California 94720, USA}
\email{whaley@berkeley.edu}

\maketitle

\begin{abstract}

For both unitary and open qubit dynamics, we compare asymmetry monotone-based bounds on the minimal time required for an initial qubit state to evolve to a final qubit state from which it is probabilistically distinguishable with fixed minimal error probability (i.e., the minimal error distinguishability time). For the case of unitary dynamics generated by a time-independent Hamiltonian, we derive a necessary and sufficient condition on two asymmetry monotones that guarantees that an arbitrary state of a two-level quantum system or a separable state of $N$ two-level quantum systems will unitarily evolve to another state from which it can be distinguished with a fixed minimal error probability $\delta \in [0,1/2]$. This condition is used to order the set of qubit states based on their distinguishability time, and to derive an optimal release time for driven two-level systems such as those that occur, e.g., in the Landau-Zener problem.  For the case of non-unitary dynamics, we compare three lower bounds to the distinguishability time, including a new type of lower bound which is formulated in terms of the asymmetry of the uniformly time-twirled initial system-plus-environment state with respect to the generator $H_{SE}$ of the Stinespring isometry corresponding to the dynamics, specifically, in terms of  $\Vert [H_{SE},\rho_{\text{av}}(\tau)]\Vert_{1}$, where $\rho_{\text{av}}(\tau):={1\over \tau}\int_{0}^{\tau}dt\, e^{-iH_{SE}t}\rho \otimes \ket{0}_{E}\bra{0}_{E} e^{iH_{SE}t}$. 
\end{abstract}

\section{Introduction}
The minimal length of time required for a given quantum state to evolve to an orthogonal state under unitary time evolution provides an ultimate bound for the processing speed of a quantum computer, regardless of the physical substrate used for the quantum information processing \cite{lloydnat}. Orthogonal states also form a valuable resource for quantum communication and for efficient quantum algorithms \cite{kitaev}. However, in practice, perfectly orthogonal states are not 
always achievable; for this reason, it is not surprising that the problem of optimally distinguishing elements of a set of nonorthogonal quantum states continues to be subject of active research (see, e.g., Refs.\cite{bergouhill,samsonov,baekwek,pirandola}) and that the resource theory of quantum coherence is profitably developed over sets of nonorthogonal, pure quantum states \cite{tanvolk}. Methods for generation and manipulation of nonorthogonal states are vital for 
high precision control of quantum dynamics and for optimal covariant quantum state estimation \cite{holevo}. The unavoidability of nonorthogonal initial and final states of realistic quantum dynamics has led to the study of generalized quantum speed limits, i.e., lower bounds on the minimal time required for an initial state to evolve into a state which is some distance from the initial state with respect to a given distance measure on state space \cite{volkoffoneparam,lloydlims}.

In this paper we consider the question of determining the minimal time $t$ required for an initial qubit state $\rho$ to evolve under unitary or non-unitary quantum dynamics $\mathcal{E}_{t}$ to a state $ \sigma \in \lbrace \mathcal{E}_{t}(\rho): t\in [0,\infty) \rbrace$ such that $\sigma$ is probabilistically distinguishable from $\rho$ with some pre-specified error $\delta$. Because the mathematical expression of distinguishability can take several forms depending on, e.g., the metric chosen for the quantum state manifold, or the choice of operational quantum state discrimination task, we define here a notion of distinguishability time which generalizes that given in Ref.\cite{volkoffoneparam} and also generalizes the notion of geometric quantum speed limit \cite{generalgeomqsl,soarespinto,deffnernjp}. 
In order to define a distinguishability time $\tau_{\delta}( \rho, \mathcal{E}_{t}, \Delta)$, representing the minimal time required for an initial quantum state $\rho$ to 
time evolve to a state from which it is distinguishable with error $\delta$, one must specify in addition to the time-evolution $\mathcal{E}_{t}$, where $\mathcal{E}_{t}$ is completely positive and trace preserving for all $t\in \mathbb{R}_{+}$, a discrimination function $\Delta$ 
that maps the set of pairs $\lbrace (\rho , \mathcal{E}_{t}(\rho))\rbrace_{t\in \mathbb{R}_{+}}$, consisting of initial state $\rho$ and its time evolved counterpart $\mathcal{E}_{t}(\rho)$, to a closed interval in $\mathbb{R}_{+}$ (positive by convention).  When $\Delta(\rho , \mathcal{E}_{t}(\rho))$ is the distance between $\rho$ and $\mathcal{E}_{t}(\rho)$ as quantified by a monotone Riemannian metric on the quantum state manifold \cite{mcp}, a lower bound for $\tau_{\delta}(\rho , \mathcal{E}_{t},\Delta)$ is considered to be a \textit{geometric quantum speed limit}. More generally, Ref.\cite{marvian} contains several examples of contractive, jointly quasiconvex functions of two quantum states that can be used to define discrimination maps. However, $\Delta$ can also be defined by a quantum information processing task aiming to distinguish $\rho$ from $\mathcal{E}_{t}(\rho)$. For example, $\Delta(\rho,\sigma)$ can be defined by $\Delta (\rho ,\sigma) := p_{\text{err}}(\rho,\sigma):= 1/2 - 1/4 \Vert \rho -\sigma \Vert_{1} \in [0,1/2]$, where the trace norm is given by $\Vert A \Vert_{1}:=\text{tr}\sqrt{A^{\dagger}A}$. The quantity $p_{\text{err}}(\rho,\sigma)$ represents the figure of merit in the task of minimal error binary quantum state discrimination \cite{helstrombook}. Other quantum state discrimination protocols such as quantum state identification \cite{hayashi,giovanettilearningstate}, which employs ancillary quantum modes (e.g., quantum training data) in lieu of perfect specification of $\rho$ in the quantum state manifold, provide alternative error functions that can be used to define $\Delta$. We emphasize that the discrimination map $\Delta$ may be asymmetric in its arguments, may depend on a given quantum state $\rho$ (see Section \ref{sec:additivity} for an example), and may be dependent on time.

From a discrimination map $\Delta$, one can define a distinguishability time $\tau_{\delta}(\rho , \mathcal{E}_{t}(\rho),\Delta)$ by \begin{equation} \tau_{\delta}( \rho, \mathcal{E}_{t}, \Delta) := \inf \lbrace t \vert \Delta(\rho, \mathcal{E}_{t}(\rho) ) = \delta \rbrace . \label{eqn:distingtime}\end{equation} Importantly, in order for $\tau_{\delta}(\rho, \mathcal{E}_{t}, \Delta )$ to have a well-defined value, there must exist a 
time $T>0$ (which may be infinite) such that $\lim_{t\rightarrow T^{-}}\Delta(\rho , \mathcal{E}_{t}(\rho)) = \delta$.  The definition in Eq.(\ref{eqn:distingtime}) can be readily generalized to the case of optimal multistate distinguishability dynamics \cite{samsonov}.
In what follows, we consider two quantum states $\rho$ and $\sigma$ to be $(1-\delta)$\textit{-distinguishable} if $\Delta(\rho,\sigma) = \delta$. 

The calculation of $\tau_{\delta}(\rho, \mathcal{E}_{t}, \Delta )$ for a given time-evolution and discrimination map may require solving for the full path $\lbrace \mathcal{E}_{t}(\rho) \vert t \ge t_{0} \rbrace$ in state space, which poses a considerable challenge for arbitrary dynamics. However, determination of the subset of values of $\delta$ such that $\tau_{\delta}( \rho, \mathcal{E}_{t}, \Delta)$ is meaningful can be obtained from, e.g., an upper bound on $\Delta (\rho , \mathcal{E}_{t}(\rho))$ for all $t$. In addition, inequalities that bound $\Delta$ by, e.g., functions of contractive distances on quantum state space, provide lower bounds on $\tau_{\delta}$ that reveal the physical properties of the state and of the dynamics that define $\tau_{\delta}$.

An outline of the paper is as follows: in Section \ref{sec:disttimesec}, we review two lower bounds on the distinguishability time when $\Delta$ is taken to be $p_{\text{err}}$ and consider these bounds in the context of unitary qubit dynamics. In Section \ref{sec:distingdynam}, we prove a necessary and sufficient condition for an initial qubit state to time evolve to a $(1-\delta)$-distinguishable state under unitary evolution generated by a time-independent Hamiltonian, and compare the distinguishability time of a two-level system to the lower bound in Eq.(\ref{eqn:mtdist}) (see Eq.(\ref{eqn:disttimemin})). Three applications of these results are subsequently discussed: 1) ordering the set of qubit states by distinguishability time (Subsection \ref{sec:faster}), calculation of a release time defined in the case of time-dependent generators of unitary qubit dynamics (Subsection \ref{sec:releasetime}), and extension of Proposition 1 to separable states of $N$ qubits (Subsection \ref{sec:additivity}). In Section \ref{sec:dissip}, we discuss three lower bounds to $\tau_{\delta}$ in the case of non-unitary evolution, and compare them in the context of a model that interpolates between Markovian and non-Markovian behavior.

\section{Minimal error distinguishability time\label{sec:disttimesec}}

Rigorous notions of uncertainty tradeoffs between measurements of energy and time have been developed in terms of the orthogonalization time, i.e., the minimum time required for an initial quantum state to evolve under the action of a given (unitary or nonunitary) quantum dynamical map to a state from which it is completely distinguishable \cite{delcampo,lloydlims,nonmarkovqsl,davidovich,dodonovrev} (see \cite{deffnerreview} for a recent review). In traditional approaches to time-energy uncertainty, a decay time or half-life of a quantum system scales inversely with the root mean square energy fluctuations of the system \cite{comminsqm,llqm}. These approaches were made mathematically rigorous by the derivation of a lower bound on the pure state orthogonalization time which scales inversely with the variance of the generator of evolution (we call this bound MT$_{\perp}$ after the seminal work of Mandelstam and Tamm \cite{mandelstam} which was subsequently put on a geometric footing by Aharanov and Anandan \cite{anandan}).

The orthogonalization time can also be bounded below by a function of the expected value of the generator of evolution (we call such a bound ML$_{\perp}$ after Margolus and Levitin \cite{margolus}). An important difference between 
the ML$_{\perp}$ and MT$_{\perp}$ bounds is that the former is an algebraic bound resulting from a linear approximation to the fidelity of the initial state and the time-evolved state, whereas the latter 
is a geometric bound which can be derived by consideration of geodesics of the Fubini-Study metric on quantum state space \cite{anandan}.

Recently, the ML$_{\perp}$ and MT$_{\perp}$ bounds have been generalized to bounds on the minimal error distinguishability time of Eq.(\ref{eqn:distingtime}) for general quantum states evolving under unitary maps \cite{volkoffoneparam}. When the unitary path is generated by $H= H^{\dagger}$ and $\Delta (\rho ,\sigma) := p_{\text{err}}(\rho,\sigma)$, the time $\tau_{\delta}$ required for $\rho$ to reach a $(1-\delta)$-distinguishable state is bounded below by the following distinguishability times \cite{volkoffoneparam}:

\begin{eqnarray}
\tau_{\delta}\ge \tau_{\delta}^{MT} &=& {2 \sin^{-1}(1-2\delta)\over \sqrt{\mathcal{F}(\rho,H)}} \label{eqn:mtdist} \\ 
\tau_{\delta}\ge \tau_{\delta}^{ML} &=& {\pi \hbar (1-\sqrt{1-(1-2\delta)^{2}})\over 2(\text{tr}(\rho H)-E_{0})}
\label{eqn:mldist}
\end{eqnarray}
where $E_{0}$ is the least eigenvalue of $H$ and $\mathcal{F}(\rho,H)$ is the quantum Fisher information on the unitary path containing $\rho$ and generated by $H$ (see Ref.\cite{helstrombook} or Section \ref{sec:distingdynam} for a definition). Clearly, the unified distinguishability bound satisfies $\lim_{\delta \rightarrow 0} \max \lbrace \tau_{\delta}^{MT}, \tau_{\delta}^{ML}\rbrace = \max \lbrace \text{MT}_{\perp} , \text{ML}_{\perp} \rbrace$. For a two-level system with time-evolution generated by $H=\hbar \omega_{0} \vec{n}\cdot \vec{\sigma}$, we will see in Section \ref{sec:distingdynam} that the states $c_{1}\ket{0}_{\vec{n}} + c_{2}\ket{1}_{\vec{n}}$ with $\vert c_{1} \vert = \vert c_{2} \vert = 1/\sqrt{2}$ are the only ones that saturate $\tau_{\delta}^{MT}$; not surprisingly, these are also the only states saturating the ML$_{\perp}$ and MT$_{\perp}$ bounds \cite{horesh,margolus}.

For a two-level system in a general state $\rho(\vec{r}) = (\mathbb{I} + \vec{r}\cdot \vec{\sigma} )/2$ evolving by the Hamiltonian $H=\hbar \omega_{0}(\vec{n}\cdot \vec{\sigma} +\mathbb{I})$ (the identity is added so that $H$ has positive semidefinite spectrum), 
application of Eq.(\ref{eqn:mldist}) yields   \begin{equation}
\tau_{\delta}^{ML} = {\pi (1-\sqrt{1-(1-2\delta)^{2}})\over 2\omega_{0}(\vec{n}\cdot \vec{r}+1)}.
\end{equation}
This bound is not consistent with the rotational symmetry of the dynamics.
For example, the state with Bloch vector $\rho(-\vec{r})$ gives a different bound than that for $\rho$. The bound $\tau_{\delta}^{ML}$ remains valid if $\vec{n}\cdot \vec{r}$ is replaced by $\vert \vec{n}\cdot \vec{r} \vert$. However, our explicit calculation of $\mathcal{F}(\rho,H)$ in Section \ref{sec:distingdynam}, combined with the fact that $\sin^{-1}(1-2\delta) \ge (\pi/2)(1-\sqrt{1-(1-2\delta)^{2}})$ for $\delta \in [0,1/2]$, leads to the conclusion that 
for $\delta \in (0,1/2)$, the Mandelstam-Tamm bound $\tau_{\delta}^{MT}$ is greater than the Margolus-Levitin bound $\tau_{\delta}^{ML}$. Hence we will focus 
here on $\tau_{\delta}^{MT}$ as 
the lower bound on the distinguishability time for the two-level system.

We note that although the derivation of $\tau^{MT}_{\delta}$ is based on combining the Bures line element $\mathcal{F}(\rho, H)$ for a unitary path generated by $H$ and the Fuchs-van de Graaf inequality  \cite{fvdg} relating the trace norm and the Bures distance (see Appendix A of \cite{volkoffoneparam}), a whole family of bounds on the distinguishability time can be constructed by combining the the line elements of other contractive metrics \cite{mcp} on quantum state space with appropriate bounds on the trace distance in terms of these contractive metrics. This leads to a family of geometric distinguishability times, analogous to the case of geometric quantum speed limits introduced in Ref.\cite{generalgeomqsl}. In our analysis of the unitary distinguishability time for a qubit in Section \ref{sec:distingdynam}, we focus on the distinguishability time defined by the Bures distance because the choice $\Delta (\rho ,\sigma) := p_{\text{err}}(\rho,\sigma)$ allows one to make use of the well known relationship between Bures distance and the trace norm given by the Fuchs-van de Graaf inequalities.

\section{Single qubit unitary distinguishability dynamics\label{sec:distingdynam}}

Before we state the main proposition, we summarize elementary results concerning minimal error distinguishability of pairs of qubit states on a path in the Bloch ball generated by a time-independent Hamiltonian. Consider a Hamiltonian $H=\hbar \omega_{0}\vec{n}\cdot \vec{\sigma}$ (where $\Vert \vec{n}\Vert = 1$ and $\vec{\sigma}:= (\sigma_{x},\sigma_{y},\sigma_{z})$) and an initial state $\ket{\psi}$. $H$ has operator norm $\hbar\omega_{0}$. By acting on $\ket{\psi}$ with the time-evolution operator $U(t):= e^{-iHt/\hbar}$ to produce $\ket{\psi(t)}$, one finds that a time $t$ such that $\langle \psi \vert \psi(t)\rangle = 0$ exists if and only if the Bloch vector $\vec{r}$ representing $\ket{\psi}$ on the Bloch sphere is orthogonal to $\vec{n}$. These states are superpositions $\ket{\phi(\varphi)} = {1\over \sqrt{2}}( \ket{0}_{\vec{n}} + e^{i\varphi}\ket{1}_{\vec{n}})$ of the lowest and highest energy states ($\ket{1}_{\vec{n}}$ and $\ket{0}_{\vec{n}}$, respectively) of $H$. Such superpositions define a great circle of states on the Bloch sphere having Bloch vector orthogonal to $\vec{n}$. A measurement of the observable $H$ in a state $\ket{\phi(\varphi)}$ has variance $1$, the largest possible value for all pure states. For any $\varphi$, the state $U(t)\ket{\phi(\varphi)}$ is orthogonal to $\ket{\phi(\varphi)}$ when $t=\pi/2$.

On the other hand, there are no completely distinguishable mixed states in the Bloch ball. Furthermore, if the initial state is mixed, it cannot be distinguished completely from any pure state. Mathematically, these facts follow from the fact that mixed states of the Bloch ball have rank two and so they cannot have support which is disjoint from the support of any other state of the Bloch ball. Hence, when the initial state is mixed, there is no hope to achieve $1$-distinguishability through any type of evolution, unitary or nonunitary. However, the evolution may still result in a $(1-\delta)$-distinguishability of initial and final states for some $\delta > 0$. Here, we consider unitary evolutions which result in $(1-\delta)$-distinguishability for $\delta >0$ and derive the set of quantum states which evolve to $(1-\delta)$-distinguishability faster than a given pure state. In the finite dimensional case considered here, the condition of $(1-\delta)$-distinguishability of two qubit states is made easier by the fact that the trace norm  $\Vert \cdot \Vert_{1}$ appearing in the expression $p_{\text{err}}(\rho,\sigma)$ can be calculated as the sum of the absolute values of the eigenvalues of $\rho -\sigma$. For the statement of Proposition 1, we again take $H= \hbar \omega_{0} \vec{n}\cdot \vec{\sigma}$. The proof is made easier by the use of a simple lemma. In this lemma, and throughout the paper, we define $\Vert \vec{a} \Vert = (\vec{a}\cdot \vec{a})^{1/2}$ to be the Euclidean norm of $\vec{a}\in \mathbb{R}^{3}$.

\textbf{Lemma}. \textit{Let the unitary path $\rho_{t} = e^{-iHt / \hbar}\rho_{0}e^{iHt / \hbar}$ generated by $H$ have initial point $\rho_{0} := {\mathbb{I}\over 2} + {\vec{r}_{0}\cdot \vec{\sigma} \over 2}$. Then the quantum Fisher information of $\rho_{t}$ is given by \begin{equation}\mathcal{F}(\rho_{t}, H) = 4\omega_{0}^{2}\Vert \vec{n} \times \vec{r}_{0} \Vert^{2} .\label{eqn:qfiform}\end{equation}
for all $t\ge 0$.}

\textit{Proof}. By definition, $\mathcal{F}(\rho_{t}, H) := \text{tr}(L^{2}\rho_{t})$, where $L=L^{\dagger}$ is the symmetric logarithmic derivative operator, i.e., the unique observable satisfying ${d\rho_{t} \over dt} = {1\over 2}(L\rho_{t} + \rho_{t} L)$ for all $t$. $L$ depends on the state $\rho_{t}$ through its Bloch vector $\vec{r}_{t}$ and also on the vector $\vec{n}$ defining $H$. It follows from ${d\rho_{t} \over dt} = -i/\hbar [H,\rho_{t}]$ that $L$ must satisfy ${1\over 2}[L,\rho_{t}]_{+} =  -i / \hbar[H,\rho_{t}] =\omega_{0} (\vec{n}\times \vec{r}_{t})\cdot \vec{\sigma} $ for all $t$. Writing $L = \vec{v}_{t}\cdot \sigma$ and solving for $\vec{v}_{t}$ results in $\vec{v}_{t}= 2\omega_{0}(\vec{n}\times \vec{r}_{t})$ so that \begin{equation} L = 2\omega_{0}(\vec{n}\times \vec{r}_{t})\cdot\vec{\sigma} . \label{eqn:symmlogderivvec}\end{equation} Taking the variance of $L$ in the state $\rho_{t}$ and using that fact that $\text{tr}(\rho L ) = 0$ gives the result $\mathcal{F}(\rho_{t},H) = 4\omega_{0}^{2}\Vert \vec{n} \times \vec{r}_{t}\Vert^{2}$. The Bloch vector $\vec{r}_{t}$ satisfies a quantum equation of motion that is the same as the classical equation of motion for a magnetic moment in a constant magnetic field which gives rise to Larmor precession, and so $\Vert \vec{n} \times \vec{r}_{t}\Vert = \Vert \vec{n} \times \vec{r}_{0}\Vert$ for all $t$. In this classical analogy, the constant value of the norm corresponds to conservation of angular momentum.
$\square$

\begin{figure}[t]
\centering
\includegraphics[scale=0.45]{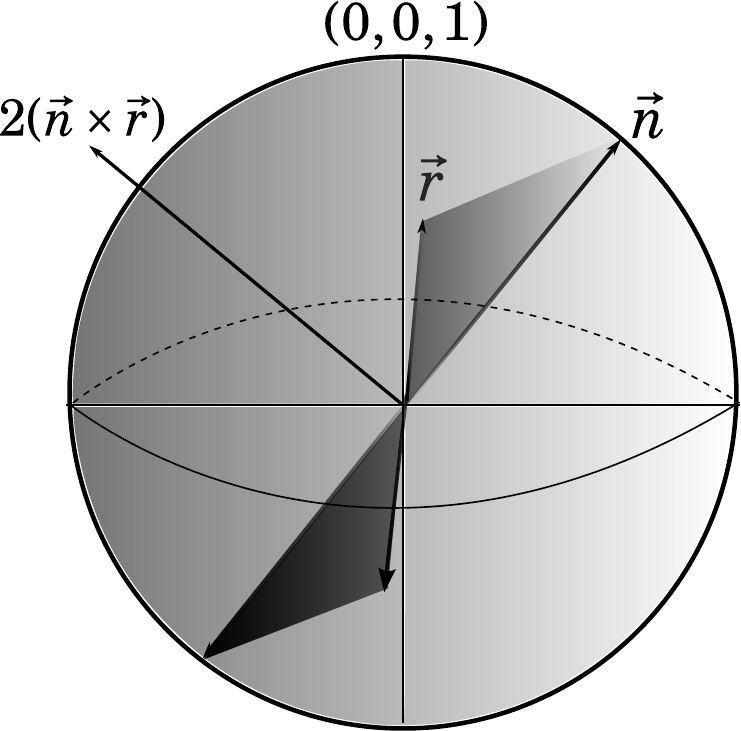}
\caption{The magnetic field vector $\vec{n}$, the Bloch vector $\vec{r}$, and the direction vector  $2( \vec{n}\times \vec{r})$  of the symmetric logarithmic derivative plotted relative to the 2-sphere. The square root of the quantum Fisher information is equal to the operator norm $\Vert L \Vert$ of $L$ which is twice the shaded area. \label{fig:mixed}}
\end{figure}

Because the observable $L$ has units of $[t]^{-1}$, $\mathcal{F}(\rho,H)$ has units of $[t]^{-2}$. The geometric relationships among the symmetric logarithmic derivative, the Hamiltonian, and the state $\rho$ are shown in Fig. \ref{fig:mixed}. Special cases of the symmetric logarithmic derivative and quantum Fisher information for a qubit evolving unitarily in the Bloch ball have been obtained previously \cite{jiang,brun,wangfisher}, but the general vectorial expression Eq.(\ref{eqn:symmlogderivvec}) provides a simple and useful formula. In the context of time-dependent quantum magnetometry with an ensemble of qubits, Eq.(\ref{eqn:qfiform}) reproduces the relevant quantum Fisher information appearing in the quantum Cram\'{e}r-Rao bound \cite{wang}.  Clearly, when $\Vert \vec{r} \Vert = 1$, i.e., when $\rho$ is pure, the quantum Fisher information takes the well known value $\mathcal{F}(\rho, H) = 4\text{tr}((\Delta H)^{2})\rho)/\hbar^{2}$ \cite{caves}.  We now state a proposition that expresses the condition for a given qubit state to $(1-\delta)$-distinguishable state under time-independent unitary dynamics.

\textbf{Proposition 1}. \textit{For a Hamiltonian $H=\hbar \omega_{0} \vec{n}\cdot \vec{\sigma}$ with $\Vert \vec{n} \Vert = 1$, there exists a $t\ge 0$ such that an initial quantum state $\rho = {1\over 2}({\mathbb{I}}+{\vec{r}\cdot \vec{\sigma}})$ evolves to a state $U(t)\rho U(t)^{\dagger}$ satisfying $p_{\text{err}}(\rho , U(t)\rho U(t)^{\dagger}) = \delta$ if and only if:} \begin{equation}
2\omega_{0}(1-2\delta)  \le \sqrt{\mathcal{F}(\rho,H)} .
\label{eqn:fisherthm}\end{equation}

\textit{Proof}. It follows from the algebra of the Pauli matrices that \begin{eqnarray}
\rho - U(t)\rho U(t)^{\dagger} &=& \sin^{2}(\omega_{0}t)(\vec{r} - (\vec{r}\cdot \vec{n})\vec{n})\cdot \vec{\sigma} \nonumber \\ &+& \sin (\omega_{0}t) \cos (\omega_{0}t )(\vec{r}\times \vec{n})\cdot \vec{\sigma}.
\label{eqn:diff}\end{eqnarray} 
We derive the conditions on $\vec{r}$ which guarantee the existence of $t$ such that $p_{\text{err}}(\rho , U(t)\rho U(t)^{\dagger})=\delta$ is satisfied. Evaluating the trace norm of Eq.(\ref{eqn:diff}) gives the following: 
\begin{eqnarray}
p_{\text{err}}(\rho , U(t)\rho U(t)^{\dagger}) &=& {1\over 2}-{1\over 2}\Vert \sin^{2}\omega_{0}t(\vec{r} - (\vec{r}\cdot \vec{n})\vec{n}) \nonumber \\ &+& \sin \omega_{0}t \cos \omega_{0}t (\vec{r}\times \vec{n}) \Vert .
\label{eqn:intermed}\end{eqnarray}
The expression in (\ref{eqn:intermed}) is equal to $\delta$ if and only if \begin{equation}
{1-2\delta \over \sqrt{\Vert \vec{r}\Vert^{2} - (\vec{r}\cdot \vec{n})^{2}}} = \vert \sin \omega_{0}t \vert .
\label{eqn:sine}
\end{equation}
Finally, a value of $t$ satisfying the above equation exists if and only if ${1-2\delta \over \sqrt{\Vert \vec{r}\Vert^{2} - (\vec{r}\cdot \vec{n})^{2}}} \le 1$. Using the Lemma to calculate the quantum Fisher information, it follows that $\sqrt{\mathcal{F}(\rho,H)} / 2\omega_{0} = \Vert \vec{n} \times \vec{r} \Vert =\sqrt{\Vert \vec{r}\Vert^{2} - (\vec{r}\cdot \vec{n})^{2}} $, which was required. $\square$

The fact that Eq.(\ref{eqn:sine}) gives an exact expression for the distinguishability time in terms of $\delta$ and the quantum Fisher information on the unitary path defined by $H$ is a consequence of the simple geometry of the Bloch ball and removes the need to analyze lower bounds for the distinguishability time for qubit states on unitary paths generated by time-independent Hamiltonians. Analogous exact expressions are not currently available even for two qubits, which are associated with the considerably more structured Lie algebra $\mathfrak{su}(4)$ of possible time-independent generators \cite{zhangwhaley}. The necessary and sufficient condition Eq.(\ref{eqn:fisherthm}) is used in Section 3.1 to order the elements of the Bloch ball based on distinguishability time and in Section 3.2 to derive an expression for the optimal release time of Landau-Zener driving in the pure state submanifold. With an appropriate choice of $\Delta$, Proposition 1 can be extended to certain unitary evolutions of multi-qubit systems prepared in separable states; this problem is discussed in Section 3.3.

Note that for a qubit undergoing unitary time evolution, the expression $\Vert [H,\rho] \Vert_{1}$, which is a faithful measure of asymmetry \cite{spekkensmarvian}, satisfies $\Vert [H,\rho] \Vert_{1} =  \sqrt{\mathcal{F}(\rho ,H)}$ and, therefore, the quantum speed limit derived in Eq.(4.1) of Ref.\cite{marvian} leads to the same necessary and sufficient condition in Proposition 1. Furthermore, the fact that $\Vert [H,\rho] \Vert_{1}$ is a faithful asymmetry monotone trivially implies the same for $\Vert [L,\rho]_{+} \Vert_{1}$. Quantum speed limits for unitary and open dynamics based on the asymmetry measure given by the Wigner-Yanase-Dyson skew information $S_{H}(\rho):= {-1\over 2}\text{tr}\left( [\sqrt{\rho},H]^{2}\right)$ were analyzed for two level quantum systems in Ref.(\cite{mondal}) and in the context of the discrimination map $\Delta(\rho , \sigma) = D_{1/2}(\rho,\sigma):= -2\log \text{tr}\sqrt{\rho}\sqrt{\sigma}$ (where $D_{1/2}(\rho,\sigma)$ denotes the $s=1/2$ quantum R\'{e}nyi relative entropy) in Ref.\cite{marvian}. In Ref.\cite{generalgeomqsl} it is proven that for unitary dynamics in the Bloch ball, the geometric quantum speed limit corresponding to the quantum Fisher information is tighter than that corresponding to the Wigner-Yanase-Dyson skew information. This conclusion carries over to the present case of distinguishability time with respect to $\Delta = p_{\text{err}}$. Specifically, in Appendix A, we derive the following bound analogous to Eq.(\ref{eqn:mtdist}):
\begin{equation}
\tau_{\delta}\ge {\sin^{-1}(1-2\delta) \over \sqrt{2 S_{H}(\rho)}} =: \tau_{\delta}^{WYD}.
\label{eqn:wydbound}
\end{equation}
For $H=\omega_{0}\vec{n}\cdot \vec{\sigma}$, $S_{H}(\rho)=\omega_{0}^{2}\left( 1 -\sqrt{1-\Vert \vec{r} \Vert^{2}} \right){\Vert \vec{r} \times \vec{n} \Vert^{2} \over \Vert \vec{r} \Vert^{2}}$ \cite{soarespinto}. Again using $\mathcal{F}(\rho , H) = 4\omega_{0}^{2} \Vert \vec{r} \times \vec{n} \Vert^{2}$, it is clear that ${\tau_{\delta}^{MT} \over \tau_{\delta}^{WYD}} \ge 1$ for all $\vec{n}$ and $\vec{r}$. It is worth pointing out the theorem of Ref.\cite{spekkensmarvian} showing that the reciprocal of any distinguishability time defined by a global discrimination map $\Delta$ and time-independent unitary $H$ is itself a measure of asymmetry with respect to $H$.

An immediate corollary of Proposition 1 is that pure qubit states are the only states for which there exists an orthogonalizing unitary evolution. For, suppose an initial state $\rho$ satisfies $\Vert \rho - \rho(t) \Vert_{1} = 2$, i.e., $\rho$ is perfectly distinguishable from $\rho(t)$. Then $\delta = 0$ in Eq.(\ref{eqn:fisherthm}) and $\mathcal{F}(\rho , H) \ge 2\omega_{0}$. This condition can be met only by a pure state proportional to $\ket{e_{+}} + e^{i\eta} \ket{e_{-}}$, where $H\ket{e_{\pm}} = \pm \hbar \omega_{0}\ket{e_{\pm}}$ and $\eta \in [0,2\pi)$.

By expanding the 
vector norm in Eq.(\ref{eqn:intermed}), it is clear that the minimal error for distinguishing $\rho$ and $\rho(t)$ occurs at time $t=\pi/2\omega_{0}$. If $\rho$ saturates the inequality (\ref{eqn:fisherthm}), then it follows from Eq.(\ref{eqn:sine})  $\rho(t=\pi /2\omega_{0})$ is $(1-\delta)$-distinguishable from $\rho$ and that $\rho(t)$ cannot be distinguished from $\rho$ with minimal error probability less than $\delta$ for any $t$.

Another implication of Eq.(10) is that the time required for an arbitrary state $\rho$ with Bloch vector $\vec{r}$ to time evolve to a $(1-\delta)$-distinguishable state (where $1-2\delta \le \Vert \vec{r} \Vert$) is given by:
\begin{eqnarray}
\tau_{\delta}(\rho , e^{-{iHt\over \hbar}}) &=& {1\over \omega_{0}}\sin^{-1}\left( { 2\omega_{0}(1-2\delta)\over \sqrt{\mathcal{F}(\rho , H)} }\right) \nonumber \\
&\ge &  {1\over \omega_{0}}\sin^{-1}\left( { (1-2\delta)\over \Vert \vec{r} \Vert }\right).
\label{eqn:disttimemin}
\end{eqnarray}
The second line gives the minimal value of $\tau_{\delta}(\rho , H)$ over all qubit Hamiltonians with operator norm $\omega_{0}$ and is obtained by taking $H$ to be a quantum brachistochrone generating a unitary evolution to the set of states that are $(1-\delta)$-distinguishable from $\rho$ \cite{carlini,mostafaz}. Note that for pure states, the lower bound $\tau_{\delta}^{MT}$ in Eq.(\ref{eqn:mtdist}) saturates the true value of $\tau_{\delta}(\rho , H)$ for the brachistochrone unitary for any $\delta$. However, the bound can be quite poor for mixed states. For example, if one takes $\tilde{\delta}$ to be the lowest minimal error probability for distinguishing between qubit states with Bloch vector magnitude $\Vert \vec{r} \Vert$ (this value is  $\tilde{\delta}={1-\Vert \vec{r} \Vert \over 2}$), and if one again takes $H=\omega_{0}{\vec{n}\cdot \vec{\sigma}}$ with $\vec{n}\cdot \vec{r} = 0$, then \begin{eqnarray} \lim_{\Vert \vec{r} \Vert \rightarrow 0 } \tau_{\tilde{\delta}}(\rho , H) - \tau_{\tilde{\delta}}^{MT} &=& \lim_{\Vert \vec{r} \Vert \rightarrow 0 }{1\over \omega_{0}}\left( {\pi \over 2} - {\sin^{-1}\Vert \vec{r} \Vert  \over \Vert \vec{r} \Vert } \right) \nonumber \\ &=& {\pi -2 \over 2\omega_{0}}. \end{eqnarray}

In Section \ref{sec:dissip}), we explore the relationship between two time-averaged asymmetry monotones and the distinguishability time $\tau_{\delta}(\rho , \mathcal{E}_{t},\Delta)$ for non-unitary dynamics $\mathcal{E}_{t}$ and discrimination map $\Delta(\rho_{1},\rho_{2}) = p_{\text{err}}(\rho_{1},\rho_{2})$.

\subsection{Qubit state ordering by distinguishability time: When are mixed states faster than pure states?\label{sec:faster}}
\begin{figure}[t]
\centering
\includegraphics[scale=0.45]{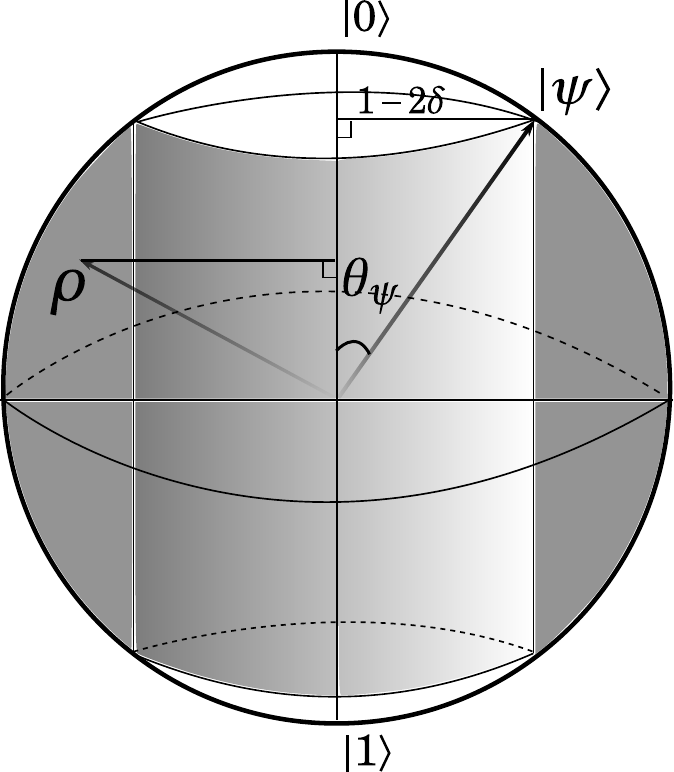}
\caption{Given a pure quantum state $\ket{\psi}$ with Bloch vector corresponding to polar angle $\theta_{\psi}$, Hamiltonian $H=\hbar \omega_{0}\sigma_{z}$, and $\delta \in [0,1/2]$, the shaded region containing $\rho$ represents those states that reach $(1-\delta)$-distinguishable states in time less than $\tau_{\delta}(\ket{\psi},e^{-i\omega_{0}t\sigma_{z}})$.\label{fig:shaded}}
\end{figure}

Having derived the necessary and sufficient condition for a given qubit state to reach a $(1-\delta)$-distinguishable state under a given unitary evolution, we are equipped to find the set of states $\sigma$ that have $\tau_{\delta}(\sigma , e^{-iHt}) < \tau_{\delta}(\rho , e^{-iHt})$ for a given state $\rho$. In particular, it follows as a corollary of Proposition 1 that there are mixed states that evolve more quickly to $(1-\delta)$-distinguishable states than certain pure states, as long as $\delta > 0$. To make this clear, we take $H=\hbar \omega_{0}\sigma_{z}$ without loss of generality and first note that the maximal quantum Fisher information of all paths in the Bloch ball generated by $H$ is achieved for the pure states $(\ket{0}+e^{i\eta}\ket{1})/\sqrt{2}$ ($\eta \in [0,2\pi)$); these are the ``fastest'' time-evolving states, reaching $(1-\delta)$-distinguishable states in time $(1/\omega_{0})\sin^{-1}(1-2\delta) \le \pi/2\omega_{0}$. Now, consider a pure state $\ket{\psi}$ with Bloch vector given by angular parameters $(\theta_{\psi} ,\varphi_{\psi})$ on the Bloch sphere with $\theta_{\psi}= \sin^{-1}(1-2\delta)$ (Fig. \ref{fig:shaded}). Then Eq.(\ref{eqn:sine}) implies that $\tau_{\delta}(\ket{\psi},e^{-i\omega_{0}t\sigma_{z}}) = \pi/2\omega_{0}$.  It follows from Proposition 1 that the set $S$ of states defined by $S=\lbrace \rho \vert \sqrt{\mathcal{F}(\rho,H)} /2 \ge \sin \theta_{\psi} =1-2\delta \rbrace$ reach $(1-\delta)$-distinguishable states and do so in a time $t \in [{1\over \omega_{0}}\sin^{-1}(1-2\delta ) , {\pi\over 2\omega_{0}}]$, i.e., in a time less than or equal to the $(1-\delta)$-distinguishability time of $\ket{\psi}$. The set of states satisfying this condition lie in the spherical ring illustrated in Fig. \ref{fig:shaded}, which is defined by polar angle $\theta > \theta_{\psi}$ and $\Vert \vec{r} \Vert \sin \theta > 1-2\delta$. 

\subsection{Release time for time dependent unitary dynamics\label{sec:releasetime}}

As an example of the utility of Proposition 1, we consider a problem concerning time dependent unitary dynamics of the set of pure qubit states. In particular, let an initial pure state $\ket{\psi(0)}$ and a time dependent Hamiltonian $H(t):= \sum_{j=x,y,z}\hbar f_{j}(t)\sigma_{j}$ be given, where $f_{j}(t): [0,\infty) \rightarrow \mathbb{R}$ are continuous functions. Examples of $H(t)$ relevant to quantum control of two-level systems include, e.g., Landau-Zener-type dynamics $H_{\text{LZ}}(t)=\sigma_{z} + f(t) \sigma_{x}$. The operator norm of $H(t)$ is a positive function of $t$ given by $\Vert H(t) \Vert = \hbar \left( \sum_{j=x,y,z}f_{j}(t)^{2} \right)^{1/2}$. The problem of interest is as follows: given $\delta \in [0,1/2]$, what is the minimal time $t_{0}\ge 0$ such that the unitary evolution $U_{t_{0}}(t):= e^{-{itH(t_{0})\over \hbar}}$ satisfies $p_{\text{err}}(\ket{\psi(t_{0})}, U_{t_{0}}(t)\ket{\psi(t_{0})}) = \delta$ for some $t>0$?

We call the time $t_{0}$ (which depends on $\delta$) the \textit{release time} for the time dependent dynamics $H(t)$. The physical picture that underlies the release time can be seen from the example of $H_{\text{LZ}}$ with the initial state $\ket{\psi(t=0)} = \ket{0}$ and a linear driving field $f(t)=t$. The experimenter would like to run the driving field $f(t)$ for as short a time $t_{0}$ as possible until a state $\ket{\psi(t_{0})}$ and Hamiltonian $H(t_{0})$ is reached such that evolution generated by the \textit{time independent} Hamiltonian $H(t_{0})$ sets $\ket{\psi(t_{0})}$ on a circular path on the Bloch sphere that contains a state that is $(1-\delta)$-distinguishable from $\ket{\psi(t_{0})}$.  

For a pure state $\rho = \rho^{2}$, the quantum Fisher information satisfies $\mathcal{F}(\rho , H) = {4\langle ( \Delta H)^{2} \rangle_{\rho} \over \hbar^{2}}$ \cite{holevo}.  For $H(t) = \sum_{j=x,y,z}\hbar f_{j}(t)\sigma_{j}$, it follows that the instantaneous quantum Fisher information $\mathcal{F}(\ket{\psi(t)},H(t))$ of the state $\ket{\psi(t)}$ with respect to the instantaneous generator $H(t)$ is given by 
\begin{eqnarray}
\mathcal{F}(\ket{\psi(t)},H(t)) &=&{4\langle ( \Delta H(t))^{2} \rangle_{\ket{\psi(t)}} \over \hbar^{2}} \nonumber \\ &=& {4\over \hbar^{2}}\left( \Vert H(t) \Vert^{2} - \langle H(t) \rangle^{2}_{\ket{\psi(t)}} \right).
\end{eqnarray}
Then, by a direct application of Proposition 1, it follows that $H(\tilde{t})$, the Hamiltonian at time $\tilde{t}\ge 0$, generates a unitary path $U_{\tilde{t}}(t)$ such that $U_{\tilde{t}}(t)\ket{\psi(\tilde{t})}$ is $(1-\delta)$-distinguishable from $\ket{\psi(\tilde{t})}$ for some $t$ if and only if
\begin{equation}
4\delta - 4\delta^{2} = { \langle \psi(\tilde{t}) \vert H(\tilde{t}) \vert \psi(\tilde{t}) \rangle^{2} \over \Vert H(\tilde{t}) \Vert^{2}} .
\label{eqn:release}
\end{equation}

The release time $t_{0}$ for $H(t)$ is given explicitly by the minimum value of $\tilde{t}$ such that Eq.(\ref{eqn:release}) is satisfied.

\subsection{Multiqubit mean minimal error distinguishability time\label{sec:additivity}}

Proposition 1 can be generalized to the setting of many qubits prepared in a separable state when the following discrimination map $\Delta$ is imposed: for separable initial state $\rho:= \sum_{j=1}^{N}p_{j}\bigotimes_{k=1}^{M}\rho^{(j)}_{k}$ and local unitary evolution $U(t):=\bigotimes_{k=1}^{M}U_{k}(t)$, where $U_{k}(t):= e^{-i \omega_{k}\vec{n}_{k}\cdot\vec{\sigma}}$ and $\sum_{j=1}^{N}p_{j}=1$, define{\footnotesize  \begin{equation} \Delta( \rho , U(t)\rho U^{\dagger}(t)) := \sum_{j=1}^{N}\sum_{k=1}^{M}{p_{j}\over M}p_{\text{err}}(\rho_{k}^{(j)}, U_{k}(t)\rho_{k}^{(j)}U_{k}^{\dagger}(t)).\label{eqn:sepdisting}\end{equation}} If only one of the $p_{j}$ is nonzero (i.e., $p_{j}= 1$ and $p_{s}=0$ for $s\neq j$), then $\Delta$ can be interpreted as the mean minimal error probability for distinguishing the single states $\rho_{k}^{(j)}$ and $U_{k}(t)\rho_{k}^{(j)}U_{k}^{\dagger}(t)$, $k=1,\ldots , M$. Therefore, for an arbitrary discrete probability distribution $p_{j}$ that defines the separable initial state $\rho$, $\Delta$ can be interpreted as the expected mean minimal error probability for distinguishing $\rho$ from $U(t)\rho U^{\dagger}(t)$ via single-site optimal measurements.  Note that the discrimination map $\Delta$ depends on the initial state $\rho$. In the case that $\omega_{j} = \omega_{k}=: \omega_{0} = \text{const.} > 0$ for all $j,k=1,\ldots ,M$, the following proposition holds.

\textbf{Proposition 2}. \textit{Let $H_{k}:=\hbar \omega_{0}\vec{n_{k}}\cdot \vec{\sigma}$ and $U(t) := \bigotimes_{k=1}^{M}e^{-{itH_{k}\over \hbar}}$. Then, with $\Delta$ and $\rho$ defined as in the paragraph above, there exists a $t\ge 0$ such that $\rho$ evolves to a state $U(t)\rho U(t)^{\dagger}$ satisfying $\Delta(\rho , U(t)\rho U(t)^{\dagger}) = \delta$ if and only if:} \begin{equation}
2\omega_{0}(1-2\delta)  \le {1\over M}\sum_{j=1}^{N}\sum_{k=1}^{M}p_{j}\sqrt{\mathcal{F}(\rho^{(j)}_{k},H_{k})} .
\label{eqn:fisherthmseo}\end{equation}

\textit{Proof}. For a given $j$ index, the mean minimal error probability  for the task of distinguishing $M$ states $\lbrace \rho_{k}^{(j)}\rbrace_{k=1,\ldots M}$, with Bloch vectors $\lbrace \vec{r}_{k}^{(j)}\rbrace_{k=1,\ldots M}$, from their corresponding images under the time evolution defined by $U(t)$ is given by
\begin{eqnarray}
p_{\text{err}}^{(j)}&:=&{1\over M}\sum_{k=1}^{M}p_{\text{err}}(\rho_{k}^{(j)},U_{j}(t)\rho_{k}^{(j)}U_{j}(t)^{\dagger}) \nonumber \\ &=&{1\over 2} - {1\over 2M}\vert \sin \omega_{0}t \vert \sum_{k=1}^{M}\Vert \vec{r}_{k}^{(j)} \times \vec{n}_{k} \Vert ,
\label{eqn:mmerror}
\end{eqnarray}
where the second line follows from Eq.(\ref{eqn:sine}). From Eq.(\ref{eqn:sepdisting}), it follows that $\Delta(\rho , U(t)\rho U(t)^{\dagger}) =\delta$ if and only if  $t$ is such that
\begin{equation}
\delta = {1\over 2}-{ \vert \sin \omega_{0}t \vert \over 2M}\sum_{j=1}^{N}\sum_{k=1}^{M}p_{j} \Vert \vec{r}_{k}^{(j)} \times \vec{n}_{k} \Vert .
\label{eqn:sepconst}
\end{equation}
A $t$ that satisfies Eq.(\ref{eqn:sepconst}) exists if and only if the condition in Eq.(\ref{eqn:fisherthmseo}) is satisfied.$\square$

The power of the necessary and sufficient condition Eq.(\ref{eqn:fisherthmseo}) becomes clear when one considers the the more general unitary evolution $U(t) := \bigotimes_{k=1}^{M}e^{-{itH_{k}\over \hbar}}$, where each $H_{k} :=\hbar \omega_{k}\vec{n_{k}}\cdot \vec{\sigma}$ is defined by its own frequency $\omega_{k}$. In this case, if we suppose that there is a time $t$ such that $\Delta(\rho , \rho(t)) = \delta$, it follows that
\begin{equation}
{1-2\delta \over {1\over M}\sum_{j=1}^{N}\sum_{k=1}^{M}{p_{j} \over 2\omega_{k}}\sqrt{\mathcal{F}(\rho^{(j)}_{k},H_{k})}} \le \vert \sin \omega_{\text{max}}t \vert
\end{equation} where $\omega_{\text{max}}:= \max_{k} \omega_{k} $. Therefore, when the local frequencies are allowed to vary, the inequality  \begin{equation}
(1-2\delta)  \le {1\over M}\sum_{j=1}^{N}\sum_{k=1}^{M}{p_{j} \over 2\omega_{k}}\sqrt{\mathcal{F}(\rho^{(j)}_{k},H_{k})},
\end{equation} analogous to Eq.(\ref{eqn:fisherthmseo}), is merely a necessary condition that such a $t$ exists.

\section{Single qubit dissipative state distinguishability dynamics\label{sec:dissip}}

In Propositions 1 and 2, we have established a necessary and sufficient condition for an arbitrary qubit state or separable multiqubit state to reach a $(1-\delta)$-distinguishable state (defined by specific discrimination maps $\Delta$) under local unitary evolution.  For the more general case of completely positive, trace preserving (CPTP) dynamics, bounds on minimal evolution time to a $(1-\delta)$-distinguishable state can be obtained from using bounds on the unitary evolution time of a larger system and the contractivity of the trace norm under the partial trace operation \cite{davidovich}. This follows from the fact that Stinespring's dilation theorem \cite{stinespring,daviesstine} 
allows one to consider any CPTP map as a unitary map acting on the system and an ancilla, followed by a partial trace over the ancillary degrees of freedom.  However, passing to the purified unitary dynamics is not necessary to derive a lower bound on $\tau_{\delta}$ for CPTP dynamics. A quantum speed limit for the Bures angle, i.e., the distinguishability time obtained by choosing the discrimination map $\Delta(\rho, \sigma)=\cos^{-1}\Vert \sqrt{\rho} \sqrt{\sigma}\Vert_{1}$, under non-unitary dynamics has been derived and studied in Ref.\cite{wisniacki,davidovich}. Along these lines, one may employ the same method as in the proof of Eq.(\ref{eqn:mtdist})  to demonstrate the direct analog of Eq.(\ref{eqn:mtdist}) for CPTP dynamics (see Appendix B):
\begin{eqnarray}
\tau_{\delta} &\ge &{2\sin^{-1}\left( 1-2\delta \right) \over {1\over \tau_{\delta}}\int_{0}^{\tau_{\delta}}dt'\,\sqrt{\mathcal{F}(\rho(t'))}}  \label{eqn:avqfi} \\ 
 &\ge& {2\sin^{-1}(1-2\delta)\over \sup_{t}\sqrt{\mathcal{F}(\rho(t))}}. \label{eqn:qfimax}
\end{eqnarray} 

Unlike the case of unitary evolution considered in Propositions 1 and 2, for the case of CPTP dynamics the quantum Fisher information is no longer a global property of the path $\rho(t)$ in the space of qubit states. Even when one considers a global quantifier of the quantum speed on a segment  $[0,T]$ of the dynamics, e.g., the time-averaged quantum Fisher information ${1\over T}\int_{0}^{T}dt \, \sqrt{\mathcal{F}(\rho(t))}$ (i.e., the mean Bures velocity), or the maximal quantum Fisher information (i.e., the maximal Bures velocity), it is not clear how such a quantifier is related to the minimal error distinguishability $p_{\text{err}}(\rho , \rho(T))$ of the initial state and final state of a segment $[0,T]$ of the dynamics.

Although the generator of generic CPTP dynamics is not given by the adjoint action $-i[H,\cdot]$ corresponding to a bounded, self-adjoint operator $H$, the time-dependent symmetric logarithmic derivative operator $L$ can still be defined for any CPTP dynamics by solving ${d\rho \over dt}={1\over 2}[\rho , L]_{+}$ for $L$. This dynamical equation allows immediate generalization of the quantum speed limit in Eq.(4.1) of Ref.\cite{marvian} to the context of CPTP dynamics. Specifically, ${d\rho \over dt} = {1\over 2}[\rho , L]_{+}$ implies $\Vert \rho(t) - \rho \Vert_{1} = \Vert \int_{0}^{t}dt \, {d\rho \over dt} \Vert_{1} \le {1\over 2}\int_{0}^{t}\Vert [\rho, L ]_{+} \Vert_{1}$. At time $\tau_{\delta}$, $\Vert \rho(\tau_{\delta}) - \rho \Vert_{1} = 2(1-2\delta)$ and, therefore,
\begin{eqnarray}
\tau_{\delta} &\ge & {2(1-2\delta) \over {1\over 2\tau_{\delta}}\int_{0}^{\tau_{\delta}}dt' \, \Vert [\rho,L]_{+} \Vert_{1}} \label{eqn:asymmmean1} \\
&\ge & {2(1-2\delta) \over {1\over 2}\sup_{t} \Vert [\rho,L]_{+} \Vert_{1}} \label{eqn:asymmmean2}
\end{eqnarray}
where both $\rho$ and $L$ are time-dependent. Note that, unlike the unitary quantum speed limit in Eq.(\ref{eqn:mtdist}), the quantum speed limit bounds (\ref{eqn:avqfi}), (\ref{eqn:asymmmean1}) require knowledge of $\tau_{\delta}$ (this is characteristic for quantum speed limits for non-unitary dynamics \cite{nonmarkovqsl,mondal,delcampo}).  A quantum speed limit for the Bures angle between a pure initial state and its time-evolved counterpart, having a form similar to Eq.(\ref{eqn:asymmmean1}), was considered as a generalization of the ML$_{\perp}$ bound to dynamics given by general CPTP dynamics in Ref.\cite{nonmarkovqsl}, but was not stated in terms of the symmetric logarithmic derivative. The same reference explores quantum speed limits for the Bures angle based on several other operator norms (including the trace norm) of ${d\rho \over dt}$.

The third lower bound for $\tau_{\delta}$ that we consider does not make use of the deviation of a path in quantum state space from a Bures geodesic (as in Eq.(\ref{eqn:avqfi})), nor the triangle inequality (as in Eq.(\ref{eqn:asymmmean1})), but rather makes direct use of the contractivity of the trace norm under CPTP dynamics. Specifically, consider the Stinespring dilation of a one-parameter CPTP dynamics $\Phi_{t}$ to be given by $\Phi_{t}(\rho) = \text{tr}_{E}\left( e^{-itH_{SE}}\rho \otimes \sigma_{E} e^{itH_{SE}}\right)$, where $\sigma_{E}$ is the initial state of the environment and where $H_{SE}$, which can be considered as the Hamiltonian of the combined system and environment, is the generator of the Stinespring isometry corresponding to $\Phi_{t}$. For a given initial state $\rho$, $\Phi_{t}$ defines the path $\lbrace \Phi_{t}(\rho) \vert \, t\ge 0 \rbrace$. For any time $T>0$, we define the uniform time-twirling of the initial state $\rho \otimes \sigma_{E}$ of the system and environment with respect to the generator of the Stinespring isometry by \begin{equation}U_{T}(\rho \otimes \sigma_{E}):= {1\over T}\int_{0}^{T}dt \,e^{-itH_{SE}}\rho \otimes \sigma_{E}\, e^{itH_{SE}}.\end{equation}  Then, by the definition of $\tau_{\delta}$, it follows that
\begin{footnotesize}
\begin{eqnarray}
2(1-2\delta) &=& \Vert \rho(\tau_{\delta}) - \rho \Vert_{1} \nonumber \\  &=& \Vert \int_{0}^{\tau_{\delta}}dt \, {d\rho \over dt} \Vert_{1} \nonumber \\ &=& \Vert \lim_{s \rightarrow 0} \int_{0}^{\tau_{\delta}}dt{\Phi_{s+t}(\rho) - \Phi_{t}(\rho) \over s}  \Vert_{1} \nonumber  \\ &=& \Vert \int_{0}^{\tau_{\delta}}dt \, \text{tr}_{E}\, [ H_{SE},e^{-itH_{SE}}\rho \otimes \sigma_{E} e^{itH_{SE}} ] \Vert_{1}  \nonumber \\ &=& \tau_{\delta}\Vert \text{tr}_{E}[H_{SE},U_{\tau_{\delta}}(\rho \otimes \sigma_{E}) ] \Vert_{1} \nonumber \\ &\le & \tau_{\delta}\Vert [H_{SE},U_{\tau_{\delta}}(\rho \otimes \sigma_{E}) ] \Vert_{1} ,
\label{eqn:stinederivation}
\end{eqnarray}
\end{footnotesize}
where we have used $\Phi_{s+t}(\rho) = \rho(t) -is\, \text{tr}_{E}[H_{SE},e^{-itH_{SE}}\rho\otimes \sigma_{E}\, e^{itH_{SE}}] + \mathcal{O}(s^{2})$ in the fourth line, and the contractivity of the trace norm in the last line. A lower bound for $\tau_{\delta}$ is given by:
\begin{equation}
\tau_{\delta}\ge {2(1-2\delta)\over \Vert [H_{SE},U_{\tau_{\delta}}(\rho \otimes \sigma_{E}) ] \Vert_{1}}.\label{eqn:stinebound}
\end{equation}

%However, within current physical realizations of quantum computers, it is not a trivial task to simply append the needed ancilla qubits and carry out the unitary map; it may therefore often be more efficient to focus efforts on engineering dissipative quantum evolutions that optimally take initial states to the desired final states in the Bloch ball.

It is worth pointing out (see also Ref.\cite{delcampo}) that the non-uniqueness of the Stinespring dilation allows an \textit{a priori} tighter lower bound improving upon Eq.(\ref{eqn:stinebound}):
\begin{equation}
\tau_{\delta}\ge {2(1-2\delta)\over \text{min}_{H_{SE},\sigma_{E}}\Vert [H_{SE},U_{\tau_{\delta}}(\rho \otimes \sigma_{E}) ] \Vert_{1}}.\label{eqn:stineboundopt}
\end{equation}
where the minimization is over the Hamiltonians and environment states that lead to the same non-unitary system dynamics. 

Finally, we note that whereas the derivations of the lower bounds (\ref{eqn:asymmmean1}) and (\ref{eqn:stinebound}) involve only a single inequality, viz., the triangle inequality for (\ref{eqn:asymmmean1}) and contractivity of the trace norm for (\ref{eqn:stinebound}), the geometric lower bound (\ref{eqn:avqfi}) involves both a Fuchs-van de Graaf inequality and the arc length inequality. As a consequence, there are two mathematical reasons why (\ref{eqn:avqfi}), or any other geometric lower bound on $\tau_{\delta}$ based on arc length with respect to a contractive metric, may fail to saturate to $\tau_{\delta}$, but only one reason why the right hand sides of (\ref{eqn:asymmmean1}) and (\ref{eqn:stinebound}) may fail to saturate to $\tau_{\delta}$. 
We put Eqs.(\ref{eqn:avqfi}), (\ref{eqn:asymmmean1}), and (\ref{eqn:stinebound}) to the test in Section \ref{sec:opentoy}, in the context of a model of amplitude damping that interpolates between Markovian and non-Markovian behavior and calculate the bound in Eq.(\ref{eqn:avqfi}) on the minimal error distinguishability time $\tau_{\delta}(\rho , \Phi_{t} , p_{\text{err}})$.

\subsection{Distinguishability dynamics of an open qubit model system \label{sec:opentoy}}

We now consider the Hamiltonian
\begin{equation}
H_{SE}:=\omega_{0}\sigma_{z}\otimes \mathbb{I}_{E} + \chi\left( \sigma_{+}\otimes \sigma_{-} + h.c. \right) ,
\end{equation} as the generator of the Stinespring isometry corresponding to the CPTP dynamics $\Phi_{t}(\rho):= \text{tr}_{E}\left( e^{-iH_{SE}t}\rho \otimes \ket{0}_{E}\bra{0}_{E} e^{iH_{SE}t} \right) $. Note that $\Vert H \Vert = \sqrt{\omega_{0}^{2}+\chi^{2}}$ and that $\Phi_{t}$ is covariant with respect to rotations about the $z$-axis, i.e., $\Phi_{t}(e^{-i\theta \sigma_{z}}\rho e^{i\theta \sigma_{z}}) = e^{-i\theta \sigma_{z}}\Phi_{t}(\rho) e^{i\theta \sigma_{z}}$. The dynamics defined by $\Phi_{t}$ allows one to consider competition between the local driving strength $\omega_{0}$ and the dissipation strength $\chi$. The Kraus form of $\Phi_{t}$ is given by $\rho(t)=\Phi_{t}(\rho) = \sum_{j=0}^{1}E_{j}(t)\rho E_{j}(t)^{\dagger}$ where
\begin{eqnarray}
E_{0}(t)&=&e^{-i\omega_{0}t}\ket{0}\bra{0} + \overline{A(t)}\ket{1}\bra{1} \nonumber \\
E_{1}(t)&=& C(t)\ket{0}\bra{1}
\label{eqn:kraus}
\end{eqnarray} 
and
\begin{eqnarray}
A(t)&:=&{\chi^{2}e^{-it\sqrt{\chi^{2}+\omega_{0}^{2}}}+(\omega_{0}-\sqrt{\chi^{2}+\omega_{0}^{2}})^{2}e^{it\sqrt{\chi^{2}+\omega_{0}^{2}}} \over 2(\chi^{2}+\omega_{0}^{2} -\omega_{0}\sqrt{\chi^{2}+\omega_{0}^{2}})} \nonumber \\
C(t)&:=& {-2i\chi\sin(t\sqrt{\chi^{2}+\omega_{0}^{2}})(\omega_{0}-\sqrt{\chi^{2}+\omega_{0}^{2}}) \over 2(\chi^{2}+\omega_{0}^{2} -\omega_{0}\sqrt{\chi^{2}+\omega_{0}^{2}})} .
\end{eqnarray}
Note that $\vert A(t) \vert^{2} + \vert C(t) \vert^{2}=1$.

The distinguishability dynamics of the channel $\Phi_{t}$ in the unitary limit $\chi \rightarrow 0$ reduces to the treatment in Section \ref{sec:distingdynam}. In particular, it follows trivially from that analysis that in the unitary limit, maximal asymmetry of the initial system-plus-environment state $\rho \otimes \ket{0}_{E}\bra{0}_{E}$ relative to the generator $H_{SE}$ of the (trivial) Stinespring isometry determines the state with the smallest value of $\tau_{\delta}$, for any $\delta$. In the following subsections, we consider whether this feature holds in the other dynamical regimes, viz., in the amplitude damping limit ($\omega_{0}=0$) and in an intermediate parameter regime, and analyze lower bounds (\ref{eqn:avqfi}), (\ref{eqn:asymmmean1}), and (\ref{eqn:stinebound}) on $\tau_{\delta}$.

\subsubsection{Amplitude damping limit ($\omega_{0}=0$)\label{sec:damplim}}
In the opposite case of amplitude damping in the absence of local unitary driving of the system (i.e., $\omega_{0}=0$), the period of the motion is $T={2\pi \over \chi}$ if $r_{1}\neq 0$ or $r_{2} \neq 0$, and is $T={\pi \over \chi}$ if $r_{1}=r_{2}=0$ and $-1\le r_{3} < 1$. The range of $\Phi_{t}$ at time $t={\pi \over 2\chi}$ is the single state $\ket{0}\bra{0}$, which is also a fixed point of the dynamics. Because all states of the Bloch ball must reach $\ket{0}\bra{0}$ at time $t={\pi \over 2\chi}$, it is clear that the mean speed of evolution is greatest for the state $\ket{1}\bra{1}$ and that the partial ordering of states by their mean speed of evolution is equivalent to their partial order based on trace distance from $\ket{0}\bra{0}$, i.e., Euclidean distance in the Bloch ball. On the time interval $[0,{\pi \over 2\chi}]$, the dynamics $\Phi_{t}$ is Markovian in the sense that no information is being transferred from the environment to the system \cite{breuermarkovian}; in particular, any two given initial states become less distinguishable as time increases from $0$ to $\pi / 2\chi$.

For an initial state $\rho$ with Bloch vector $(r_{1},0,r_{3})$, the Bloch vector of $\Phi_{t}(\rho)=\rho(t)$ is given by $\vec{r}(t)= \left( r_{1}\cos \chi t , 0, r_{3} + (1-r_{3})\sin^{2}\chi t \right)$. For any time $t$, the minimal error probability for binary distinguishability of $\rho$ and $\rho(t)$ is given by
\begin{eqnarray}
p_{\text{err}}(\rho , \rho(t)) &=& {1\over 2} -{1\over 4} \left( r_{1}^{2}\left( 1-\cos \chi t \right)^{2} \right. \nonumber \\ 
&+& \left. \left( r_{3}-1\right)^{2}\sin^{4}\chi t \right)^{1/2} .
\end{eqnarray}
Therefore, we see that the set of quantum states that orthogonalize under the dynamics consists of the union of the following two sets: 1) pure states with $r_{3}=0$ (orthogonalize in time $\pi/\chi$), and 2) $\ket{1}\bra{1}$ (orthogonalizes in time $\pi / 2\chi$; it is the fastest state, and the only state that orthogonalizes within the time interval $[0,\pi/2\chi ]$ for which the dynamics is Markovian).

\begin{figure}[t]
\centering
\includegraphics[scale=0.35]{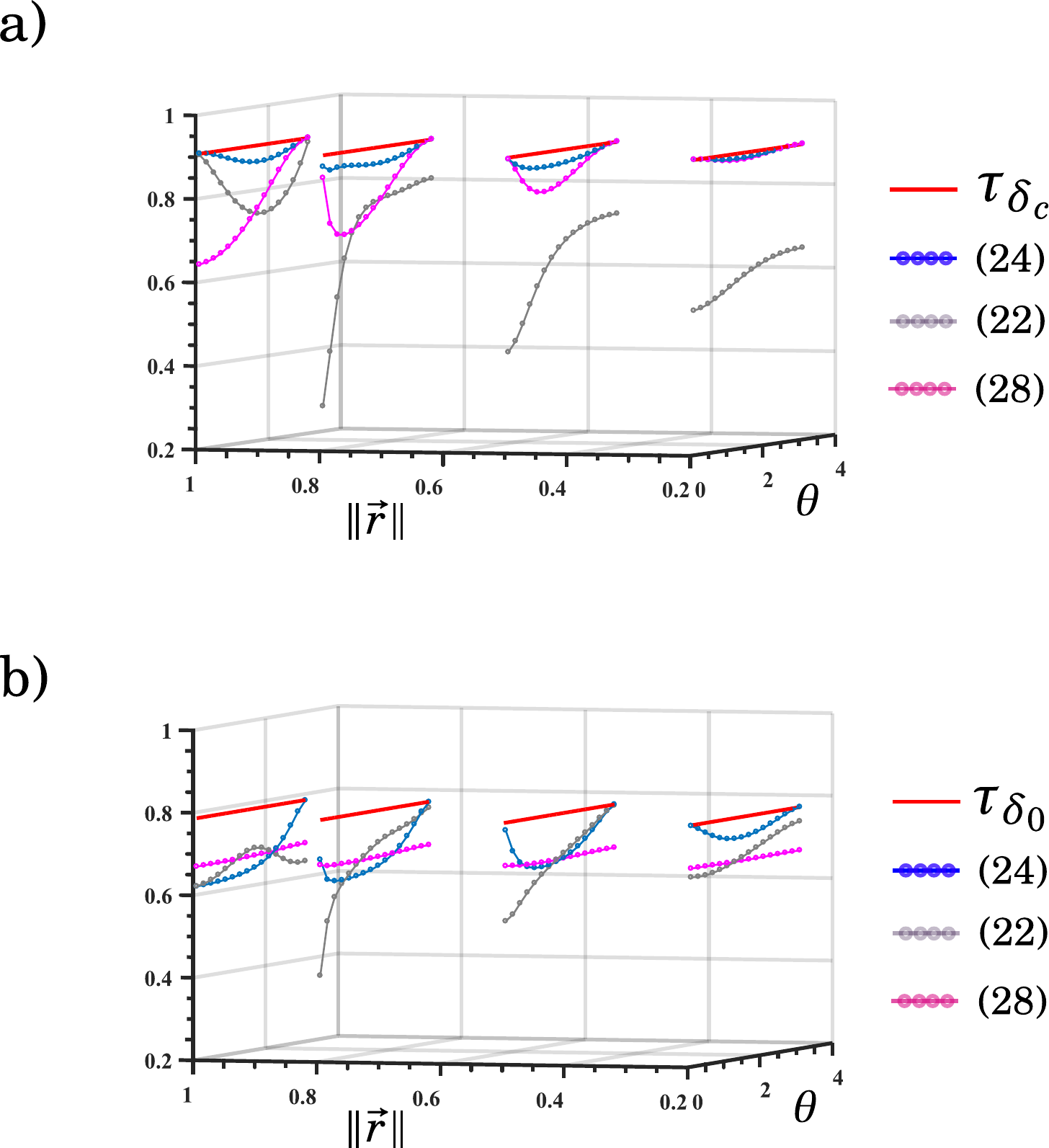}
\caption{a) Bounds (\ref{eqn:avqfi}), (\ref{eqn:asymmmean1}), and (\ref{eqn:stinebound}) on $\tau_{\delta}$ in the case of amplitude damping. b) Bounds (\ref{eqn:avqfi}), (\ref{eqn:asymmmean1}), and (\ref{eqn:stinebound}) in the intermediate case. Values of $\Vert \vec{r} \Vert$ are $0.2$, $0.5$, $0.8$, $1.0$. For each bound, the relevant time average is approximated by the discrete mean on the interval $[0,\tau]$ ($\tau = \tau_{c}$ or $\tau = \tau_{0}$) with time step $0.002$. \label{fig:comp}}
\end{figure}

We now compare the distinguishability time bounds Eqs.(\ref{eqn:avqfi}), (\ref{eqn:asymmmean1}), and (\ref{eqn:stinebound}). For a given initial state $\rho = {1\over 2}\left( \mathbb{I} + (r_{1},0,r_{3})\cdot \vec{\sigma}\right)$, consider the error probability $\delta_{c}:= p_{\text{err}}(\rho , \rho(\pi/2\chi ) )={1\over 2}-{1\over 4}\sqrt{ r_{1}^{2}+(1-r_{3})^{2}}$. For any initial state $\rho$, $\delta_{c}$  is the minimal error that is achievable for binary discrimination of $\rho$ and $\rho(t)$ over all $t\in [0,\pi / 2\chi ]$. Furthermore, $t=\pi / 2\chi$ is the unique value of $t\in [0,\pi / 2\chi]$ which gives $p_{\text{err}}(\rho , \rho(t) ) = \delta_{c}$. Therefore, we compare the bounds (\ref{eqn:avqfi}), (\ref{eqn:asymmmean1}), and (\ref{eqn:stinebound}) by substituting $\tau_{\delta_{c}} = \pi / 2\chi $ for $\tau_{\delta}$ and determining which of the resulting lower bounds better approximates the value $\pi / 2\chi$ from below. In Fig. \ref{fig:comp}a), we see that for initial states near $\ket{1}\bra{1}$ or for mixed states with $\Vert \vec{r} \Vert \lesssim 0.8$, both bounds (\ref{eqn:asymmmean1}) and (\ref{eqn:stinebound}) are tighter than the bound (\ref{eqn:avqfi}).  Therefore, the $\omega_{0}=0$ regime of the present model serves as an example of Markovian evolution for which the bound (\ref{eqn:asymmmean1}) is globally tighter than the geometric bound based on the mean quantum Fisher information (\ref{eqn:avqfi}) or the asymmetry of the time-twirled state with respect to the Stinespring isometry generator (\ref{eqn:stinebound}).

\subsubsection{Intermediate regime ($\omega_{0}\neq 0$, $\chi \neq 0$)}
\begin{figure}[t]
\centering
\includegraphics[scale=0.4]{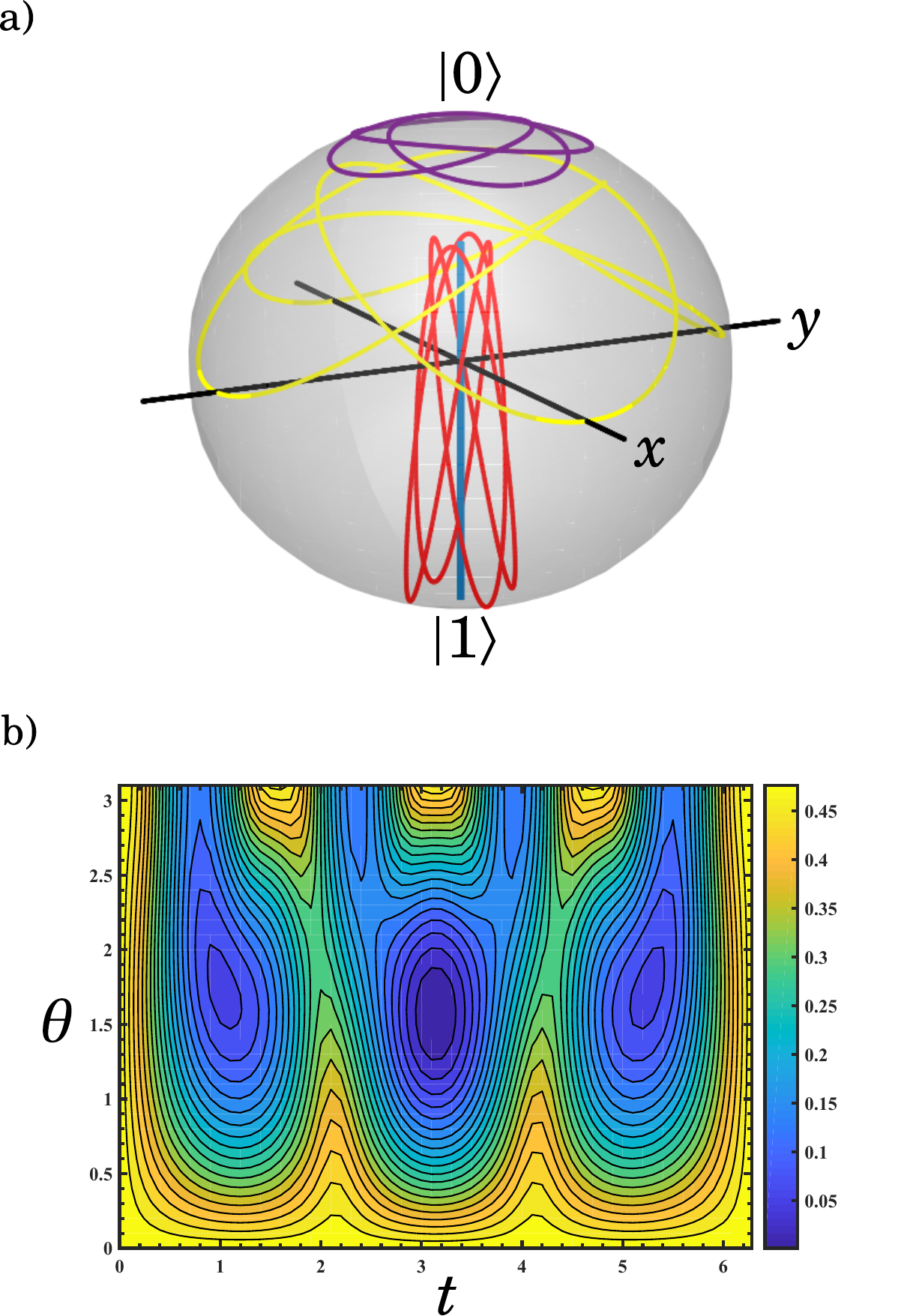}
\caption{a) Bloch vectors of $\rho(t)$ on a single period of the dynamics $\Phi_{t}$ with initial Bloch vectors: $(0,0,-1)$ (blue), $(\sin ({14\pi \over 15}),0,\cos ({14\pi \over 15}))$ (red), $(1,0,0)$ (yellow), $({1\over 2},0,{\sqrt{3}\over 2})$ (purple). b) Minimal error probability $p_{\text{err}}(\rho,\rho(t))$ for initial pure states $\rho$ with Bloch vector $(\sin \theta , 0 ,\cos \theta)$. Note the reduction of the period by $1/4$ when the initial state is $\ket{1}\bra{1}$. For all trajectories, $\chi = \sqrt{3}$, $\omega_{0}=1$.\label{fig:blochpaths}}
\end{figure}

In the case that $\omega_{0}\neq 0$, there does not exist a time such that the range of the channel $\Phi_{t}$ is the single pure state $\ket{0}\bra{0}$. Furthermore, it follows from Eq.(\ref{eqn:kraus}) that if $\omega_{0}\neq 0$, and if the Bloch vector of the initial state $\rho$ satisfies $r_{1}=r_{2}=0$, then $\Phi_{t}$ is periodic with period $T={\pi \over \sqrt{\chi^{2} + \omega_{0}^{2}}}$. If $\omega_{0} \neq 0 $ and if $\rho$ has a Bloch vector with non-zero $r_{1}$ or $r_{2}$, then the motion is periodic if and only if $\omega_{0}/\sqrt{\chi^{2}+\omega_{0}^{2}} \in \mathbb{Q}$. In the latter case, if one denotes $ \omega_{0}/\sqrt{\chi^{2}+\omega_{0}^{2}} = {q/p}$ with $(p,q) \in \mathbb{N}$ a pair of coprime natural numbers, then the period is $T={2\pi q \over \omega_{0}}$. Because of the rotation covariance of the dynamics, we consider only initial states $\rho = {\mathbb{I}+(r_{1},0,r_{3})\cdot \vec{\sigma} \over 2}$, for which the Bloch vector of $\rho(t)$ takes the form 
\begin{eqnarray}r_{1}(t)&=&\text{Re}(A(t)e^{-i\omega_{0}t})r_{1}  \nonumber \\ r_{2}(t)&=& -\text{Im}(A(t)e^{-i\omega_{0}t})r_{1} \nonumber \\ r_{3}(t)&=& (1-\vert A(t)\vert^{2}) +\vert A(t) \vert^{2}r_{3}.\label{eqn:tdepbloch}\end{eqnarray} In Fig. \ref{fig:blochpaths}a), several trajectories are shown. It follows from Eq.(\ref{eqn:tdepbloch}) that a pure state that satisfies $r_{3}=0$ orthogonalizes in time $\pi q /\omega_{0}$ (i.e., a half-period of its dynamics). Let us compare this orthogonalization time to the orthogonalization time $\pi/ \chi$ in the amplitude damping regime. For a qubit state initialized in $(1/\sqrt{2})(\ket{0}+\ket{1})$ and for the parameters $\omega_{0}=1$, $\chi = \sqrt{3}$, $q$ takes the value 1. Therefore, since $\pi/ \chi < \pi$, the orthogonalization cycle is longer in this intermediate regime despite the fact that the local system driving has been increased. Increasing the ratio of $\omega_{0}$ to $\chi$ does not necessarily change the situation. For example, if $\chi = 3$, the orthogonalization of $(1/\sqrt{2})(\ket{0}+\ket{1})$ occurs in time $\pi / 3$ in the amplitude damping regime, while turning on $\omega_{0}$ to $\omega_{0}=4$ (so that $q/p = \omega_{0}/\sqrt{\omega_{0}^{2} + \chi^{2}} = 4/5 \Rightarrow q=4$) changes the orthogonalization cycle period to $\pi$.

Note that if one considers $\delta \neq 0$, i.e., considers a nonzero minimal error probability for distinguishing $\rho$ and $\rho(t)$, the dynamical regime which is considered to allow ``fast'' time evolution can change. For example, in Section \ref{sec:damplim}, we saw that in the Markovian time interval $[0,\pi / 2\chi]$ for the regime defined by $\omega_{0}=0$, $\chi \neq 0$, the state $(1/\sqrt{2})(\ket{0}+\ket{1})$ evolves to $\ket{0}$, from which it can be distinguished with minimal error $(\sqrt{2}-1) / 2\sqrt{2}$. For $\chi = \sqrt{3}$, note that $\pi/2\chi \approx 0.91$. For the case of $\omega_{0}=1$, $\chi = \sqrt{3}$ in the present dynamical regime, $(1/\sqrt{2})(\ket{0}+\ket{1})$ reaches a state from which it can be distinguished with minimal error probability $(\sqrt{2}-1) / 2\sqrt{2}$ in time $t \approx 0.66$. To sum up, for faster orthogonalization, one prefers $\omega_{0}=0$, $\chi = \sqrt{3}$; for faster semi-orthogonalization, one prefers $\omega_{0}=1$, $\chi = \sqrt{3}$.

In order to compare the bounds (\ref{eqn:avqfi}), (\ref{eqn:asymmmean1}), and (\ref{eqn:stinebound}) in the present dynamical regime, we consider the error probability $\delta_{0}:= p_{\text{err}}(\rho, \rho(t_{0}))$, where $t_{0}$ is the time at which the $r_{3}$ component of the Bloch vector reaches its maximal value. The value of $t_{0}$ is readily computable from $r_{3}(t)$ in Eq.(\ref{eqn:tdepbloch}), and is given by $t_{0}={\pi \over 2\sqrt{ \omega_{0}^{2} + \chi^{2}}}$. For concreteness, we present calculations performed for $\chi = \sqrt{3}$, $\omega_{0}=1$, so that $t_{0}=\pi/4$. On the interval $t\in [0,\pi/4]$, $p_{\text{err}}(\rho , \rho(t) )$ is monotonically decreasing for all initial states $\rho$. Therefore, $\tau_{\delta_{0}} = t_{0} = \pi/4$. Note that $\delta_{0}$ is given by

\begin{eqnarray}
\delta_{0} &=&{1\over 2}- {1\over 4}\left( \vert C(t_{0}) \vert^{4} (r_{3}-1)^{2} \right. \nonumber \\ & +& \left. \vert A(t_{0})e^{-i\omega_{0}t_{0}} -1 \vert^{2}r_{1}^{2} \right)^{1/2} .
\end{eqnarray}
Analogously to the approach used for the amplitude damping regime, we compare the bounds (\ref{eqn:avqfi}), (\ref{eqn:asymmmean1}), and (\ref{eqn:stinebound}) by setting ${\delta}={\delta_{0}}$ and determining which right hand side of (\ref{eqn:avqfi}), (\ref{eqn:asymmmean1}), or (\ref{eqn:stinebound}) better approximates the value $\pi /4$ from below. In Fig. \ref{fig:comp}b), one sees that (\ref{eqn:asymmmean1}) remains the tightest lower bound deep in the Bloch ball. However, the geometric lower bound (\ref{eqn:avqfi}) and bound (\ref{eqn:stinebound}) dominate the bound (\ref{eqn:asymmmean1}) in specific regions of the Bloch ball. 

In the present dynamical regime, the combined lower bound given by the supremum of the right hands sides of (\ref{eqn:avqfi}), (\ref{eqn:asymmmean1}), and (\ref{eqn:stinebound}) is, on average, comparatively less tight to the distinguishability time $\tau_{\delta_{0}}$ than the combined lower bound in the Markovian regime is to $\tau_{\delta_{c}}$. We consider this to be due to the fact that the lower bounds (\ref{eqn:avqfi}), (\ref{eqn:asymmmean1}), and (\ref{eqn:stinebound}) take into account only the asymmetry properties of the system, and not the asymmetry of the environment.

\section{Conclusion}
We have introduced the formal notion of distinguishability time with respect to a discrimination map $\Delta$ and quantum dynamics $\mathcal{E}_{t}$ as a generalization of the orthogonalization time. By defining $\Delta$ to coincide with the minimal error probability $p_{\text{err}}$ in the task of distinguishing a quantum state from its time-evolved image, we solved for the distinguishability time of time-independent unitary evolution of a two-level system (Eq.(\ref{eqn:sine})) and of homogeneous time-independent unitary evolution of a separable state of $N$ two-level systems. This enabled a determination of the set of qubit states that evolve to $(1-\delta)$-distinguishable states faster than a given qubit state under unitary time-evolution. In the case of time-dependent unitary evolution generated by a controllable driving Hamiltonian $H(t)$, we derived a condition for the release time, viz., the earliest time point $t_{0}$ at which the driving can be halted such that a given qubit state evolves to a $(1-\delta)$-distinguishable state under further time-independent unitary evolution $U_{t_{0}}(t)=e^{-itH(t_{0})/\hbar}$.

For the case of non-unitary evolution, we compared three lower bounds on $\tau_{\delta}$ in a model that interpolates between Markovian and non-Markovian behavior. In the Markovian time interval of the amplitude damping regime ($\omega_{0}=0$), we found that the bound (\ref{eqn:asymmmean1}) is tight to the distinguishability time. In contrast, in the non-Markovian intermediate regime ($\omega_{0}\neq 0$, $\chi\neq 0$), the bound (\ref{eqn:avqfi}) and the new bound (\ref{eqn:stinebound}) approximate the distinguishability time more closely than (\ref{eqn:asymmmean1}) for certain angles and levels of state mixedness.

Although the connections between the speed of quantum evolution and Markovian and non-Markovian character of the dynamics are beginning to be understood \cite{adessononmarkov}, general theorems are still lacking. For a specific model of non-unitary evolution, we showed that the state transformation $(1/\sqrt{2})(\ket{0} + \ket{1}) \mapsto \ket{0}$ proceeds faster in a non-Markovian regime than in a Markovian regime. However, if one aims to implement the state transformation $(1/\sqrt{2})(\ket{0} + \ket{1}) \mapsto (1/\sqrt{2})(\ket{0} - \ket{1}) $ at high frequency, increasing the system driving may in fact increase the cycle period. We expect the present work to provide a basis for future investigations of the relation between distinguishability and general quantum dynamics of one- and few-qubit systems.

\acknowledgments

This work was supported by National Science Foundation Grant No. CHE-1213141. T.J.V. is supported by the Korea Research Fellowship Program through the National Research Foundation of Korea (NRF) funded by the Ministry of Science and ICT (2016H1D3A1908876) and by the Basic Science Research Program through the NRF funded by the Ministry of Education (2015R1D1A1A09056745).

\appendix
\section{Distinguishability time bound based on Wigner-Yanase-Dyson skew information}

We prove the bound appearing in Eq.(\ref{eqn:wydbound}). Consider the following inequalities:
\begin{eqnarray}
{1\over 4}\Vert \rho - \rho(t) \Vert_{1}^{2} &\le & 1- \left( \text{tr} \rho^{1/2}\rho(t)^{1/2} \right)^{2} \nonumber \\  &\le & 1-\cos^{2}\left(t S_{H}(\rho) \right)
\end{eqnarray} 
where the first line is derived in Ref.\cite{holevoquasiequiv} and the second line follows from the inequality $t\sqrt{ - \text{tr}\left( [\sqrt{\rho},H]^{2}\right)} \ge \cos^{-1}\text{tr}\rho^{1/2}\rho(t)^{1/2}$ \cite{mondal} which holds for $t$ such that $t\sqrt{ - \text{tr}\left( [\sqrt{\rho},H]^{2}\right)} \in [0,\pi/2]$. By definition, ${1\over 2}\Vert \rho - \rho(\tau_{\delta}) \Vert_{1} = 1-2\delta$. Therefore,
\begin{equation}
\sin \left( \tau_{\delta}\sqrt{ - \text{tr}\left( [\sqrt{\rho},H]^{2}\right)} \right) \ge 1-2\delta
\end{equation}
which implies Eq.(\ref{eqn:wydbound}).

\section{Distinguishability time bound based on mean quantum Fisher information}

We prove the lower bound in Eq.(\ref{eqn:avqfi}) for the distinguishability time under arbitrary quantum dynamics. Note the following inequalities:
\begin{eqnarray}
{1\over 4}\Vert \rho - \rho(t) \Vert_{1}^{2} &\le& 1-\left( \text{tr}\sqrt{ \sqrt{\rho}\rho(t)\sqrt{\rho}} \right)^{2} \nonumber \\ &\le & 1-\cos^{2}\left( {1\over 2}\int_{0}^{t}dt' \mathcal{F}(\rho(t')) \right),
\label{eqn:app2}
\end{eqnarray}
where the first inequality is a Fuchs-van de Graaf inequality and the second inequality follows from the fact that the length of a path in quantum state space (with respect to the Bures arc length) is at least the Bures angle, i.e.,  ${1\over 2}\int_{0}^{t}dt' \mathcal{F}(\rho(t'))\ge \cos^{-1}  \text{tr}\sqrt{ \sqrt{\rho}\rho(t)\sqrt{\rho}}$. Since, by definition, $\Vert \rho - \rho(\tau_{\delta}) \Vert_{1} = 2(1-2\delta)$, Eq.(\ref{eqn:avqfi}) follows by rearrangement of Eq.(\ref{eqn:app2}).

\bibliographystyle{naturemag}
\bibliography{disting_refs_update}

\end{document}